\documentclass[a4paper]{revtex4}
\usepackage{graphicx}
\graphicspath{{figures/}}
\usepackage{fancyhdr}
\usepackage{amsmath}
\newcommand{\GeV}{\mathrm{GeV}}

\usepackage{color}

\begin{document}

\title{Higgs-Portal Dark Matter for GeV Gamma-Ray Excess}

%
\author{P. Ko}
\affiliation{Korea Institute for Advanced Study, Seoul, South Korea}
\author{Wan-Il Park}
\affiliation{Departament de Fisica Teorica and IFIC, Universitat de Valencia-CSIC,
E-46100, Burjassot, Spain}
\author{Yong Tang}
\affiliation{Korea Institute for Advanced Study, Seoul, South Korea}

\begin{abstract}
We present Higgs-Portal dark matter (DM) models to explain the reported Galactic Center GeV gamma-ray excess. Naive effective theories are inconsistent with direct detection constraint for the relevant parameter range. Simple extended models with dark gauge symmetries can easily accommodate the gamma-ray excess through the Higgs-Portal coupling while satisfying various constraints. 
\end{abstract}

\maketitle

\thispagestyle{fancy}


\section{Introduction}
Recently, it was reported in \cite{Daylan:2014rsa} (see also~\cite{other}) that a possible GeV-scale gamma-ray excess might be due to DM annihilation into $b\bar{b}$ with canonical cross section. 
The most discussed Higgs-portal DM models are the following effectively described theories,
\begin{align}
&\Delta {\cal L}_S = -{1\over 2} m_S^2 S^2  -  {1\over 2} \lambda_{hSS}  H^\dagger H  S^2 \;,  \\
&\Delta {\cal L}_f = - m_\chi \bar \chi \chi - 
{\lambda_{h\chi\chi}\over \Lambda} H^\dagger H \bar \chi \chi \;, \\
&\Delta {\cal L}_X = {1\over 2} m_X^2 X_\mu X^\mu + {1\over 2} \lambda_{hXX}  H^\dagger H X_\mu X^\mu \;, 
\end{align}
for spin 0,1/2, and 1, respectively. In the above Lagrangian, the only relevant new degree of freedom is the dark matter field and all other fields have been assumed heavy and integrated out. Here, we have neglected the kinetic terms and possible self-interaction terms for DM as well. 

\begin{figure}[b]
\includegraphics[scale=0.5]{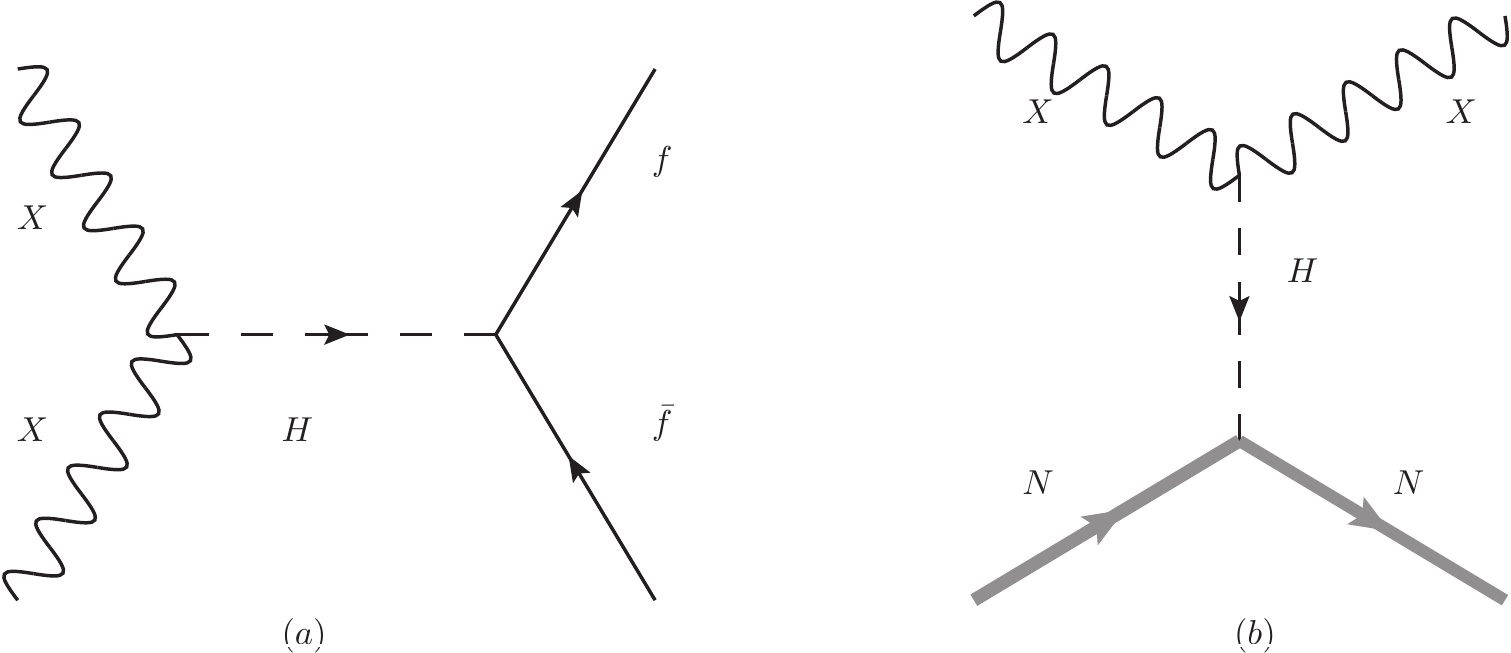}
\caption{Feynman diagrams for annihilation and direct detection.\label{fig:bbnn}}
\end{figure}

However, the reported gamma-ray excess is usually explained through the annihilation channel, Fig.~\ref{fig:bbnn}(a). This would naturally lead to large scattering cross section for direct detection in Fig.~\ref{fig:bbnn}(b).  For fermionic DM, $\lambda_{h\chi\chi}$-term would give p-wave suppressed cross section, which then leads to a much larger cross section during the freeze-out time in the early Universe. A possible term, $H^\dagger H \bar \chi \gamma_5 \chi$, is not discussed here. 

We consider a vector dark matter, $X_{\mu}$, which is associated
with dark gauge symmetry, taking $U(1)_{X}$ as an example. The simplest model will
be the one including only a complex scalar $\Phi$, whose vacuum expectation
value (vev) is responsible for the mass of $X_{\mu}$:
\begin{eqnarray}
{\cal L} & = & -\frac{1}{4}X_{\mu\nu}X^{\mu\nu}+(D_{\mu}\Phi)^{\dagger}(D^{\mu}\Phi)-\lambda_{\Phi}\left(\Phi^{\dagger}\Phi-\frac{v_{\Phi}^{2}}{2}\right)^{2}\nonumber \\
 &  & -\lambda_{H\Phi}\left(H^{\dagger}H-\frac{v_{H}^{2}}{2}\right)\left(\Phi^{\dagger}\Phi-\frac{v_{\Phi}^{2}}{2}\right)-\lambda_{H}\left(H^{\dagger}H-\frac{v_{H}^{2}}{2}\right)^{2}+\mathcal{L}_{\mathrm{SM}}.\label{eq:full_theory}
\end{eqnarray}
We here neglected the kinetic mixing term $X_{\mu\nu}B^{\mu\nu}$, but for non-abelian groups, such term does not appear. More details can be found in \cite{Ko:2014gha} and other phenomenologies is also discussed in \cite{Baek:2014goa}.

The above covariant derivative $D_{\mu}$ on $\Phi$ is defined as
\[
D_{\mu}\Phi=(\partial_{\mu}-ig_{X}X_{\mu})\Phi.
\]
Assuming that the $U(1)_{X}$-charged complex scalar $\Phi$ develops
a nonzero vev, $v_{\Phi}$, 
\[
\Phi=\frac{1}{\sqrt{2}}\left(v_{\Phi}+\varphi\right),
\]
which breaks $U(1)_{X}$ spontaneously. Therefore $X_{\mu}$ gets mass $M_{X}=g_{X}v_{\Phi}$,
and  dark Higgs field $\varphi$ will mix with the SM Higgs field $h$ through the Higgs-portal $\lambda_{H\Phi}$ term. The mixing matrix $O$ between the two scalar
fields is defined as 
\begin{equation}
\left(\begin{array}{c}
h\\
\varphi
\end{array}\right)=\left(\begin{array}{cc}
c_{\alpha} & s_{\alpha}\\
-s_{\alpha} & c_{\alpha}
\end{array}\right)\left(\begin{array}{c}
H_{1}\\
H_{2}
\end{array}\right),
\end{equation}
 where $s_{\alpha}(c_{\alpha})\equiv\sin\alpha(\cos\alpha)$, $H_{i}(i=1,2)$
are the mass eigenstates with masses $m_{i}$. $H_{1}$ is conventionally identifid
as the new observed Higgs at the LHC with $m_1=125$GeV. The mixing angle $\alpha$ is determined by
\begin{eqnarray*}
\sin2\alpha=\frac{2\lambda_{H\Phi}v_{H}v_{\Phi}}{m_{2}^{2}-m_{1}^{2}}.
\end{eqnarray*}

The small mixing between $H_2$ and $H_1$ enable $H_2$ decay into SM fermion pairs  which give the required prompt gamma-ray flux since heavy quarks can be dominant channels. 

\begin{figure}[b]
\includegraphics[scale=0.9]{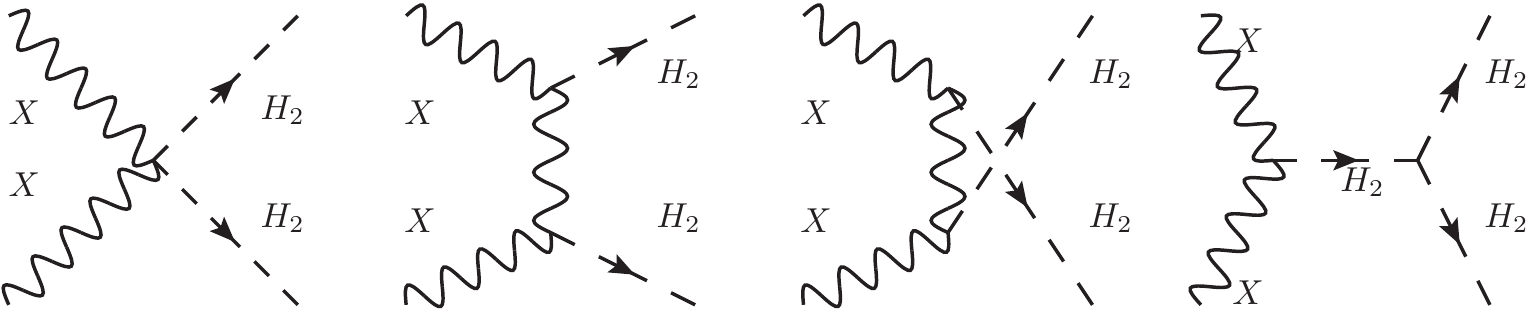}
\caption{The dominant annihilation channels. $H_2$ can decay into DM fermions due to its mixing with $H_1$. }
\end{figure}
\section{Higgs-Portal DM Model with Local $Z_3$ Symmetry}\label{sec:modelz3}
\begin{figure}[t]
\includegraphics[width=0.9\textwidth, height=0.16\textwidth]{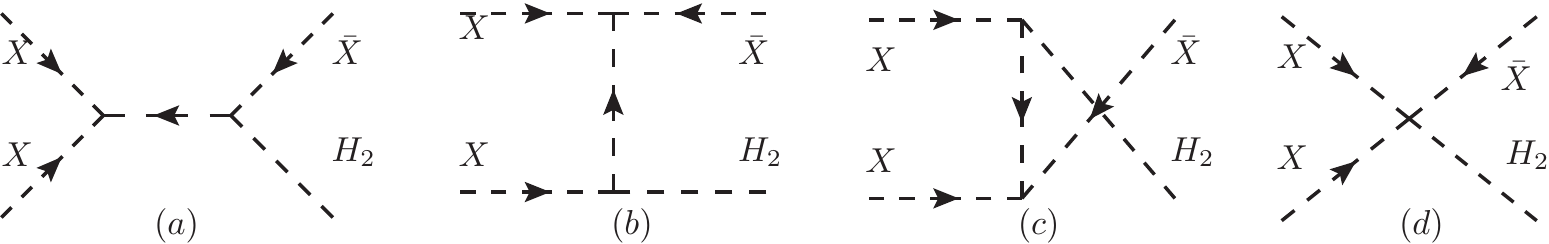}
\includegraphics[width=0.7\textwidth, height=0.16\textwidth]{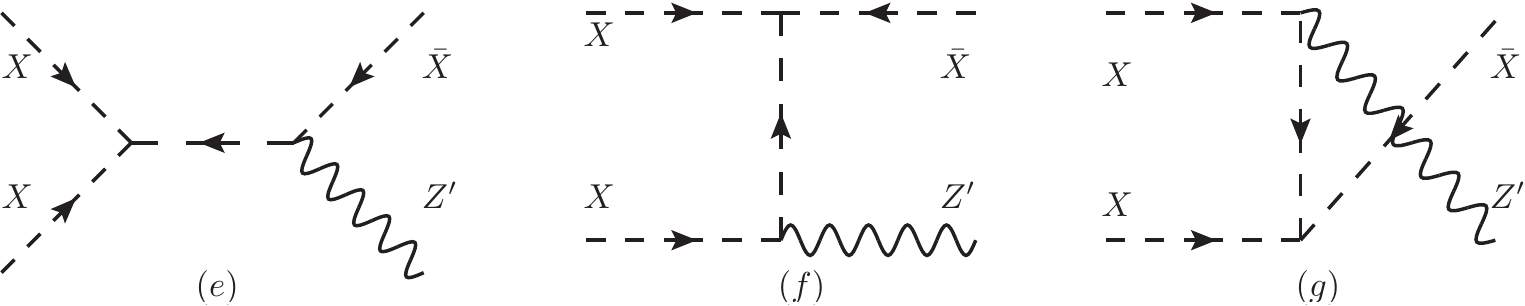} 
\caption{Feynman diagrams for $XX$ semi-annihilation into $H_2$ and $Z'$. 
\label{fig:semi-annihilation}}
\end{figure}

In this section, we consider a scalar DM model with local $Z_3$ symmetry~\cite{localz3} in which a local $U(1)_{X}$ symmetry is spontaneously broken into $Z_{3}$. This can be realized with two complex scalar fields, $\phi_{X}$ and $X$, 
with the $U(1)_{X}$ charges equal to $1$ and $1/3$, respectively. Here we intend to explain the gamma-ray signal~\cite{Z3gamma}. The renormalizable Lagrangian is
\begin{eqnarray}
{\cal L} & = & {\cal L}_{{\rm SM}}-\frac{1}{4}\tilde{X}_{\mu\nu}\tilde{X}^{\mu\nu}-\frac{1}{2}\sin\epsilon\tilde{X}_{\mu\nu}\tilde{B}^{\mu\nu}+D_{\mu}\phi_{X}^{\dagger}D^{\mu}\phi_{X}+D_{\mu}X^{\dagger}D^{\mu}X-V,\nonumber \\
V & = & -\mu_{H}^{2}H^{\dagger}H+\lambda_{H}\left(H^{\dagger}H\right)^{2}-\mu_{\phi}^{2}\phi_{X}^{\dagger}\phi_{X}+\lambda_{\phi}\left(\phi_{X}^{\dagger}\phi_{X}\right)^{2}+\mu_{X}^{2}X^{\dagger}X+\lambda_{X}\left(X^{\dagger}X\right)^{2}\nonumber \\
 &  & {}+\lambda_{\phi H}\phi_{X}^{\dagger}\phi_{X}H^{\dagger}H+\lambda_{\phi X}X^{\dagger}X\phi_{X}^{\dagger}\phi_{X}+\lambda_{HX}X^{\dagger}XH^{\dagger}H+
 \left( \lambda_{3}X^{3}\phi_{X}^{\dagger}+H.c. \right) ,  \label{eq:potential}
\end{eqnarray}
where  $D_{\mu}\equiv\partial_{\mu}-i\tilde{g}_{X}Q_{X}\widetilde{X}_{\mu}$. 
The coupling $\lambda_{3}$ is a real, positive number.

After the symmetry breaking, we have
\begin{eqnarray}
\langle H\rangle=\frac{1}{\sqrt{2}}\left(\begin{array}{c}
0\\
v_{h}
\end{array}\right),\;\langle\phi_{X}\rangle=\frac{v_{\phi}}{\sqrt{2}},\;\langle X\rangle=0,\label{eq:vacuumstate}
\end{eqnarray}
where $H$ and $\phi_{X}$ have non-zero vacuum expectation values (VEVs). Then  
EW symmetry is broken into $U(1)_{\rm em}$ and dark $U(1)_{X}$ gauge symmetry is 
broken into discrete $Z_3$, which stabilizes scalar DM $X$.  Expand scalar fields around Eq.~(\ref{eq:vacuumstate}), 
\begin{equation}
H\rightarrow\frac{v_{h}+h}{\sqrt{2}},\;\phi_{X}\rightarrow\frac{v_{\phi}+\phi}{\sqrt{2}},\; X\rightarrow\frac{x}{\sqrt{2}}e^{\mathrm{i}\theta}\textrm{ or }\frac{1}{\sqrt{2}}\left(X_{R}+iX_{I}\right),
\end{equation}
we find two scalar bosons $h$ and $\phi$ mix with each other, resulting in two mass 
eigenstates $H_{1}$ and $H_{2}$ with
\begin{equation}
\left(\begin{array}{c}
H_{1}\\
H_{2}
\end{array}\right)=\left(\begin{array}{cc}
\cos{\alpha} & {}-\sin{\alpha}\\
\sin{\alpha} & \cos{\alpha}
\end{array}\right)\left(\begin{array}{c}
h\\
\phi
\end{array}\right),
\end{equation}
in terms of the mixing angle $\alpha$. We shall identify $H_1$ as the recent discovered Higgs boson with $M_{H_1}\simeq 125\GeV$ and treat $M_{H_2}$ freely. EW and dark gauge symmetry breaking also leads to mixing among neutral gauge bosons. The mass eigenstates $( A_\mu , Z_\mu , Z_\mu^{'} )$ are defined by
\begin{equation}
\left(\begin{array}{c}
\tilde{B}_{\mu}\\
\tilde{W}_{3\mu}\\
\tilde{X}_{\mu}
\end{array}\right)=\left(\begin{array}{ccc}
c_{\tilde{W}} & -\left(t_{\epsilon}s_{\xi}+s_{\tilde{W}}c_{\xi}\right) & s_{\tilde{W}}s_{\xi}-t_{\epsilon}c_{\xi}\\
s_{\tilde{W}} & c_{\tilde{W}}c_{\xi} & -c_{\tilde{W}}s_{\xi}\\
0 & s_{\xi}/c_{\epsilon} & c_{\xi}/c_{\epsilon}
\end{array}\right)\left(\begin{array}{c}
A_{\mu}\\
Z_{\mu}\\
Z'_{\mu}
\end{array}\right).\label{eq:mixing2}
\end{equation}
New parameters are introduced for parametrization:  
\begin{eqnarray}
 &  & c_{\tilde{W}}\equiv\cos\theta_{\tilde{W}}=\frac{g_{2}}{\sqrt{g_{1}^{2}+g_{2}^{2}}},\;\tan2\xi=-\frac{m_{\tilde{Z}}^{2}s_{\tilde{W}}\sin2\epsilon}{m_{\tilde{X}}^{2}-m_{\tilde{Z}}^{2}\left(c_{\epsilon}^{2}-s_{\epsilon}^{2}s_{\tilde{W}}^{2}\right)},\nonumber \\
 &  & t_{x}\equiv\tan{x},\; c_{x}\equiv\cos{x}\;\mathrm{and}\; s_{x}\equiv\sin{x}\;\mathrm{for}\; x=\epsilon,\xi,\nonumber \\
 &  &m_{\tilde{X}}^{2}=\hat{g}_{X}^{2}v_{\phi}^{2},\;\hat{g}_{X}=\tilde{g}_X/c_\epsilon,\; m_{\tilde{Z}}^{2}=\frac{1}{4}\left(g_{1}^{2}+g_{2}^{2}\right)v_{h}^{2}.
\end{eqnarray}
From Eq.~(\ref{eq:mixing2}) we notice that SM particles can have interaction with dark photon $Z'_{\mu}$ and $Z'_{\mu}$ can decay into SM fermion pairs. The physical masses for four gauge bosons at tree level in our model are given by 
\begin{align}
m_{A}^{2} =  0,\;   & m_{W}^{2}  =  m_{\tilde{W}}^{2}=\frac{1}{4}g_{2}^{2}v_{h}^{2},  \\
m_{Z}^{2} =  m_{\tilde{Z}}^{2}\left(1+s_{\tilde{W}}t_{\xi}t_{\epsilon}\right),\;
& m_{Z'}^{2}  =  \frac{m_{\tilde{X}}^{2}}{c_{\epsilon}^{2}\left(1+s_{\tilde{W}}t_{\xi}t_{\epsilon}\right)}.
\end{align} 

The gamma ray from DM annihilation is given by
\begin{equation}\label{eq:flux}
\frac{d^2 \Phi}{dE_\gamma d\Omega}=\frac{1}{8\pi}\sum_f\frac{\langle \sigma v\rangle^f_\textrm{ann}}{ M_{\textrm{DM}}^2}\frac{d N^f_\gamma }{dE_\gamma}\int_0^\infty dr \rho ^2 \left(r'\left( r,\theta \right) \right),
\end{equation}
where $\langle \sigma v\rangle^f_\textrm{ann}$ is the thermal annihilation cross section, $d N^f_\gamma/dE_\gamma$ is prompt gamma-ray spectrum, $r'=\sqrt{r_\odot^2 + r^2 -2r_\odot  r \cos \theta}$, $r$ is the distance to earth from the DM annihilation point,  $r_\odot\simeq 8.5$kpc for solar system and $\theta$ is the observation angle between the line-of-sight and the center of Milky Way. We use NFW density profile for DM,
\begin{equation}\label{eq:haloprofile}
\rho\left(r\right)=\rho_\odot \left[\frac{r_\odot}{r}\right]^\gamma \left[\frac{1+r_\odot/r_c}{1+r/r_c}\right]^{3-\gamma},
\end{equation}
with parameters $r_c\simeq 20$kpc and $\rho_\odot \simeq 0.4\textrm{GeV}/\textrm{cm}^3$.

\begin{figure}[t]
\includegraphics[width=0.47\textwidth, height=0.42\textwidth]{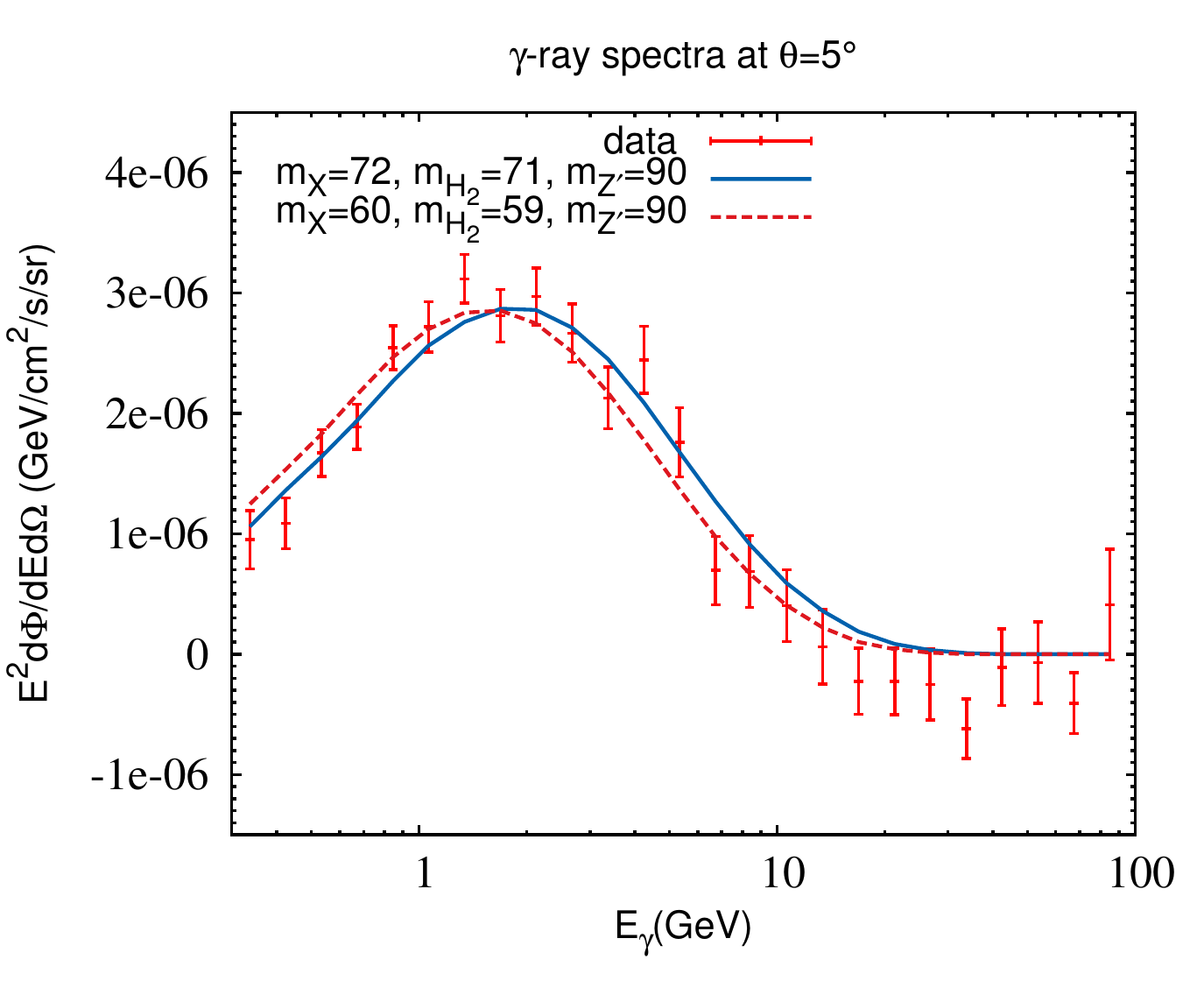}
\includegraphics[width=0.47\textwidth, height=0.42\textwidth]{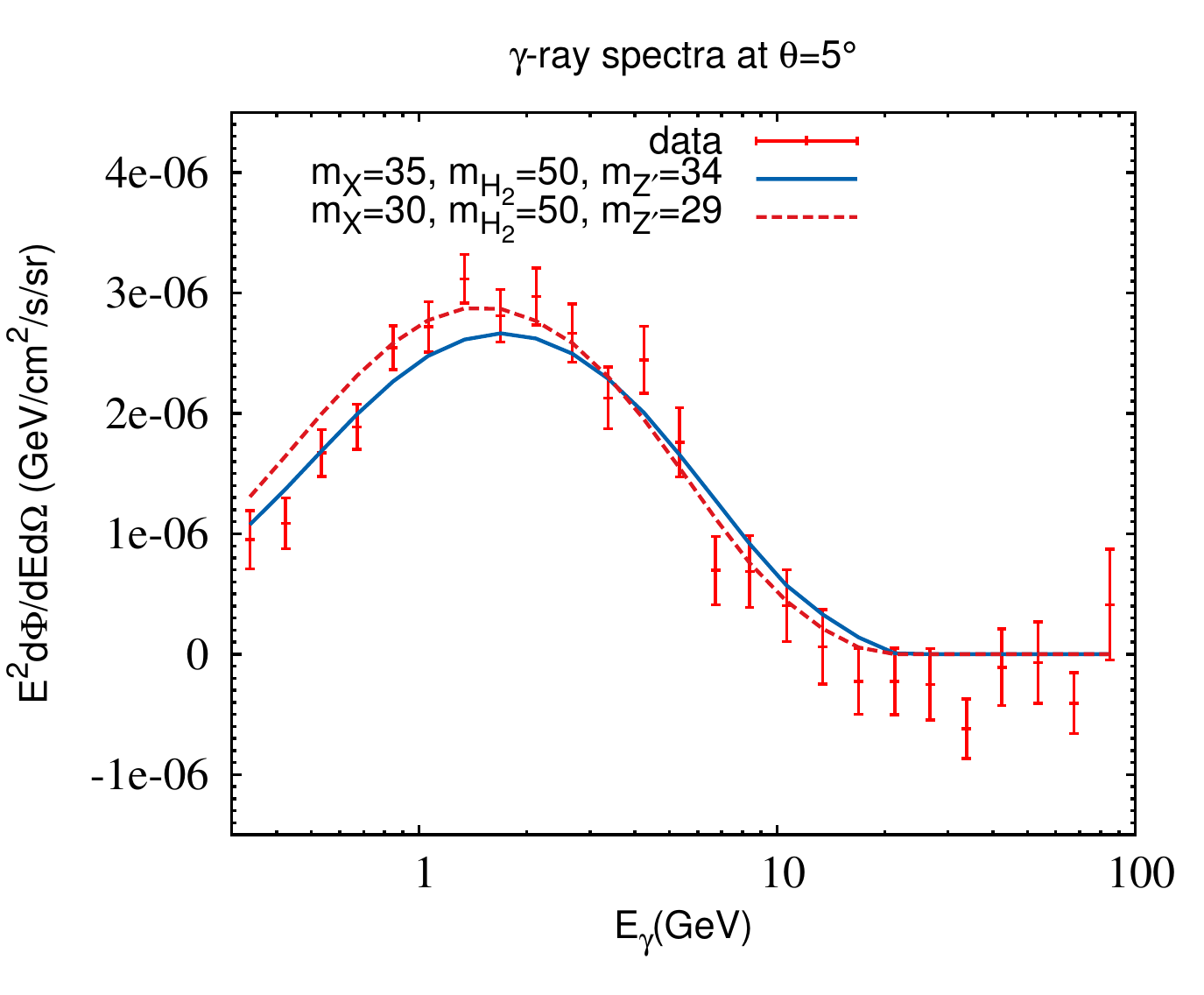} 
\caption{Gamma-ray spectra from X's (semi-)annihilation into $H_2$(left) and $Z'$(right). $m_{H_2}$ or $m_{Z'}$ in GeV is chosen to be close to $m_X$ to avoid large Lorentz boost. Data points are extracted from \cite{Daylan:2014rsa}.
\label{fig:gammaspectrum}}
\end{figure}

We have chosen several illustrating examples in Fig.~\ref{fig:gammaspectrum}, which shown that the gamma-ray spectra from DM annihilation in our $Z_3$ model can indeed fit with the reported excess. Both annihilation and semi-annihilation process are relevant.  

\section{Summary}
We have shown Higgs-portal dark matter models with hidden symmetries can explain the recently reported GeV gamma-ray from Galactic center. Depending on the specific model setup, Higgs-portal interaction or kinetic mixing term can induce the dark higgs or $Z'$'s decay into SM fermions, which then result in the required gamma-ray flux. Details of parameter space can be found in~\cite{Ko:2015ioa}.

\begin{acknowledgments}
This work is supported in part by National Research Foundation of Korea (NRF) Research 
Grant 2012R1A2A1A01006053. 
\end{acknowledgments}

\bigskip 

\end{document}